\documentclass{article}\usepackage[]{graphicx}\usepackage[]{xcolor}
\makeatletter
\def\maxwidth{ %
  \ifdim\Gin@nat@width>\linewidth
    \linewidth
  \else
    \Gin@nat@width
  \fi
}
\makeatother

\definecolor{fgcolor}{rgb}{0.345, 0.345, 0.345}

\usepackage{framed}
\makeatletter
 {\par\unskip\endMakeFramed%
 \at@end@of@kframe}
\makeatother

\definecolor{shadecolor}{rgb}{.97, .97, .97}
\definecolor{messagecolor}{rgb}{0, 0, 0}
\definecolor{warningcolor}{rgb}{1, 0, 1}
\definecolor{errorcolor}{rgb}{1, 0, 0}
\newenvironment{knitrout}{}{} % an empty environment to be redefined in TeX

\usepackage{alltt}

\usepackage{arxiv}

\usepackage[utf8]{inputenc} % allow utf-8 input
\usepackage[T1]{fontenc}    % use 8-bit T1 fonts
\usepackage{hyperref}       % hyperlinks
\usepackage{url}            % simple URL typesetting
\usepackage{booktabs}       % professional-quality tables
\usepackage{amsfonts}       % blackboard math symbols
\usepackage{nicefrac}       % compact symbols for 1/2, etc.
\usepackage{microtype}      % microtypography
\usepackage{graphicx}
\usepackage{natbib}
\usepackage{doi}

\usepackage{lipsum}

\usepackage{longtable}
\usepackage{colortbl}
\usepackage{amsmath}
\usepackage{caption}
\usepackage{multirow}
\hypersetup{hidelinks}
\usepackage{geometry}
\usepackage{pdflscape}
\usepackage{bm}
\usepackage{algorithm, algcompatible, algpseudocode}
\usepackage{eqparbox}

\DeclareMathOperator*{\argmax}{argmax}

\newdimen{\algindent}
\algnewcommand\LeftComment[2]{%
\hspace{#1\algindent}$\triangleright$ \eqparbox{COMMENT}{#2} \hfill %
}

\algnewcommand{\algorithmicgoto}{\textbf{go to}}%
\algnewcommand{\Goto}[1]{\algorithmicgoto~\ref{#1}}%

\DeclareCaptionLabelFormat{AppendixTables}{A.#2}

\definecolor{lightgray}{rgb}{0.83, 0.83, 0.83}

\newcommand{\dataset}{{\cal D}}

\newcommand{\ie}{\textit{i.e.}}
\newcommand{\eg}{\textit{e.g.}}

\newcommand{\bstat}{\widehat{\text{BS}}(t)}
\newcommand{\bsbar}{\mathcal{\widehat{BS}}(t_1, t_2)}
\newcommand{\bskap}{\mathcal{\widehat{BS}}_0(t_1, t_2)}

\newcommand{\ntest}{N_{\text{test}}}

\newcommand{\secref}[1]{Section \ref{#1}}

\newcommand{\tabref}[1]{Table \ref{#1}}
\newcommand{\tabrefAppendix}[1]{Table A.\ref{#1}}

\newcommand{\nope}[1]{}

\title{Accelerated and interpretable oblique random survival forests}

\author{ \href{https://orcid.org/0000-0001-7399-2299}{\includegraphics[scale=0.06]{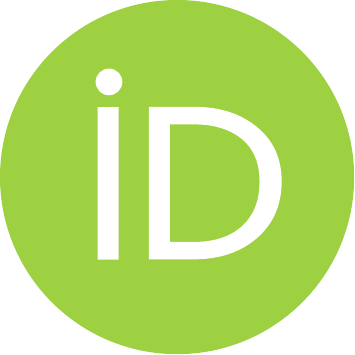}\hspace{1mm}Byron C. Jaeger} \\
	Department of Biostatistics and Data Science\\
	Wake Forest University School of Medicine\\
	Winston-Salem, NC 27157, USA \\
	\texttt{bjaeger@wakehealth.edu}
	\AND
   Sawyer Welden \\
   Department of Biostatistics and Data Science\\
   Wake Forest University School of Medicine\\
   Winston-Salem, NC 27157, USA \\
   \texttt{swelden@wakehealth.edu}
   \And
   Kristin Lenoir \\
   Department of Biostatistics and Data Science\\
   Wake Forest University School of Medicine\\
   Winston-Salem, NC 27157, USA \\
   \texttt{klenoir@wakehealth.edu}
	 \And
	 Jaime L.~Speiser \\
	 Department of Biostatistics and Data Science\\
	 Wake Forest University School of Medicine\\
	 Winston-Salem, NC 27157, USA \\
	 \texttt{jspeiser@wakehealth.edu}
	 \And
	 Matthew Segar \\
   Department of Cardiology \\
   Texas Heart Institute \\
   Houston, TX 77030, USA \\
	 \texttt{Matthew.Segar@BCM.edu}
   \And
	 Ambarish Pandey \\
   Division of Cardiology, Department of Internal Medicine \\
   University of Texas Southwestern Medical Center \\
   Dallas, TX 75235, USA \\
	 \texttt{Ambarish.Pandey@UTSouthwestern.edu}
	 \And
	 Nicholas M.~Pajewski \\
	 Department of Biostatistics and Data Science\\
	 Wake Forest University School of Medicine\\
	 Winston-Salem, NC 27157, USA \\
	 \texttt{npajewsk@wakehealth.edu}
}

\renewcommand{\shorttitle}{\textit{arXiv} Accelerated \& interpretable oblique RSFs}

\hypersetup{
pdftitle={Accelerated and interpretable oblique random survival forests},
pdfsubject={q-bio.NC, q-bio.QM},
pdfauthor={Byron C.~Jaeger},
pdfkeywords={Random Forest, Survival, Efficiency, Variable Importance},
}
\IfFileExists{upquote.sty}{\usepackage{upquote}}{}
\begin{document}

\maketitle

\begin{abstract}

	The oblique random survival forest (RSF) is an ensemble supervised learning method for right-censored outcomes. Trees in the oblique RSF are grown using linear combinations of predictors to create branches, whereas in the standard RSF, a single predictor is used. Oblique RSF ensembles often have higher prediction accuracy than standard RSF ensembles. However, assessing all possible linear combinations of predictors induces significant computational overhead that limits applications to large-scale data sets. In addition, few methods have been developed for interpretation of oblique RSF ensembles, and they remain more difficult to interpret compared to their axis-based counterparts. In this article, we introduce and evaluate a method to increase computational efficiency of the oblique RSF and a method to estimate importance of individual predictor variables with the oblique RSF. Our strategy to reduce computational overhead makes use of Newton-Raphson scoring, a classical optimization technique that we apply to the Cox partial likelihood function within each non-leaf node of decision trees. We estimate the importance of individual predictors for the oblique RSF by negating each coefficient used for the given predictor in linear combinations, and then computing the reduction in out-of-bag accuracy. In general benchmarking experiments, we find that our implementation of the oblique RSF is approximately 450 times faster with equivalent discrimination and superior Brier score compared to existing software for oblique RSFs. We find in simulation studies that `negation importance' discriminates between relevant and irrelevant predictors more reliably than permutation importance, Shapley additive explanations, and a previously introduced technique to measure variable importance with oblique RSFs based on analysis of variance. All methods pertaining to oblique RSFs in the current study are available in the \texttt{aorsf} R package.

\end{abstract}

% keywords can be removed
\keywords{Random Forests \and Survival \and Efficiency \and Variable Importance}

\section{Introduction}

Risk prediction may reduce the burden of disease by guiding strategies for prevention and treatment in a wide range of domains \citep{moons2012riskII, moons2012riskI}. The random survival forest (RSF; \citet{ishwaran2008random, hothorn2006unbiased}) is a supervised learning algorithm that has been used frequently for risk prediction \citep{wang2017selective}. Similar to random forests (RFs) for classification and regression \citep{breiman2001random}, The RSF is a large set of de-correlated and randomized decision trees, with each tree contributing to the ensemble's prediction function. Notable characteristics of the RSF include uniform convergence of its ensemble survival prediction function to the true survival function, first shown by \citet{ishwaran2010consistency} and later by \citet{cui2017consistency} under more general conditions. However, \citet{cui2017consistency} noted that the RSF is at a disadvantage when predictors are correlated and some are not relevant to the censored outcome, which is a strong possibility when large medical databases are leveraged for risk prediction.

A potential approach to improve the RSF when predictors are correlated and some are not relevant to the censored outcome is to use oblique trees instead of axis based trees. Axis based trees split data using a single predictor, creating decision boundaries that are perpendicular or parallel to axes of the predictor space \citep[see][Chapter~2]{breiman2017classification}. Oblique trees split data using a linear combination of predictors, creating decision boundaries that are neither parallel nor perpendicular to axes of their contributing predictors \citep[see][Chapter~5]{breiman2017classification}. \citet{menze2011oblique} examined prediction accuracy of RFs in the presence of correlated predictors and found that oblique RFs had substantially higher prediction accuracy compared to axis-based RFs. Similarly, \citet{jaeger2019oblique} found that growing RSFs with oblique rather than axis-based survival trees reduced the RSF's concordance error, with improvements ranging from 2.5\% to 24.9\% depending on the data analyzed.

Oblique trees have at least two notable drawbacks compared to axis-based trees. First, finding a locally optimal oblique decision rule may require exponentially more computation than an axis-based rule. If $p$ predictors are potentially used to split $n$ observations, up to $\mathcal{O}(n^p)$ oblique splits can be assessed versus $\mathcal{O}(n \cdot p)$ axis-based splits \citep{heath1993induction, murthy1994system}. Second, estimating variable importance (VI) using permutation (a standard method for RFs) may be less effective in ensembles of oblique trees, as permuting the values of one predictor may not destabilize decisions that are based on linear combinations of predictors. Although VI is one of the most widely used strategies to interpret random forests \citep{ishwaran2019standard}, few studies have investigated VI for oblique random forests \citep[see][Section~5]{menze2011oblique}, and fewer have investigated VI specifically for the oblique RSF.

The rest of this article is organized as follows. \secref{sec:background} reviews prior studies that have developed methods related to those introduced in the current study. In \secref{sec:aorsf}, we reduce the computational cost of oblique RSFs (\ie, accelerate them) with a scalable algorithm to identify linear combinations of coefficients. In \secref{sec:negation_vi}, we improve the interpretability of oblique RSFs with `negation VI', a method to estimate VI that flips the sign of coefficients in linear combinations of predictors instead of permuting predictor values. We evaluate these methods with general benchmarking experiments and simulation studies in \secref{sec:numeric}. In \secref{sec:discussion}, we summarize results from the current study and present ideas connecting the current work to existing frameworks and methods for RSFs that future studies may engage with. The accelerated oblique RSF and multiple methods to compute VI for oblique RSFs are available in the \texttt{aorsf} R Package.

% However, there remains considerable potential to improve the RSF in settings where training samples are not large enough to guarantee asymptotic properties or do not align with the conditions required for consistency of the RSF.

% Random forests (RFs) are large sets of de-correlated, randomized decision trees \citep{breiman2001random}, which can be axis based or oblique.Although using oblique rather than axis-based trees to grow a RSF may improve its prediction accuracy,

\section{Related work} \label{sec:background}

Sections \ref{sec:rw_forests} and \ref{sec:rw_vi} briefly summarize prior studies that have developed methods related to the oblique RSF and VI, respectively.

\subsection{Axis-based and oblique random forests} \label{sec:rw_forests}

After \citet{breiman2001random} introduced the axis-based and oblique RF, numerous methods were developed to grow oblique RFs for classification or regression tasks \citep{menze2011oblique, zhang2014oblique, rainforth2015canonical, zhu2015reinforcement, poona2016investigating, qiu2017oblique, tomita2020sparse, katuwal2020heterogeneous}. However, oblique splitting approaches for classification or regression may not generalize to censored outcomes \citep[\eg, see][Section~4.5.1]{zhu2013tree}, and most research involving the RSF has focused on forests with axis-based trees \citep{wang2017selective}.

Building on prior research for bagging survival trees \citep{hothorn2004bagging}, \citet{hothorn2006unbiased} developed an axis-based RSF in their framework for unbiased recursive partitioning, more commonly referred to as the conditional inference forest (CIF). \citet{zhou2016random} developed a rotation forest based on the CIF and \citet{wang2017random} developed a method for extending the predictor space of the CIF. \citet{ishwaran2008random} developed an axis-based RSF with strict adherence to the rules for growing trees proposed in \citet{breiman2001random}.  \citet{jaeger2019oblique} developed the oblique RSF following the bootstrapping approach described in Breiman's original RF and incorporating early stopping rules from the CIF.

Fast algorithms to fit axis-based RSFs are available in the \texttt{randomForestSRC} R package \citep{randomForestSRC} and the \texttt{ranger} \citep{ranger} R package. \texttt{randomForestSRC} provides a unified interface to grow RFs in a wide range of analyses, and \texttt{ranger} is designed to grow RFs efficiently using high dimensional data. Fast algorithms to fit the CIF are provided by the \texttt{party} R package \citep{hothorn2010party}, which provides a computational toolbox for recursive partitioning using conditional inference trees. \citet{jaeger2019oblique} developed the \texttt{obliqueRSF} package and found it was approximately 30 times slower than \texttt{party} and nearly 200 times slower than \texttt{randomForestSRC}. Few studies have developed software with fast algorithms for oblique RSFs that have comparable speed compared to algorithms for axis-based RSFs.

\subsection{Variable importance} \label{sec:rw_vi}

Several techniques to estimate VI have been developed since \citet{breiman2001random} introduced permutation VI, which is defined for each predictor as the difference in a RF's estimated prediction error before versus after the predictor's values are randomly permuted. \citet{strobl2007bias} identified bias in permutation VI driven by variable selection bias and effects induced by bootstrap sampling, and proposed an unbiased permutation VI measure based on unbiased recursive partitioning \citep{hothorn2006unbiased}. \citet{menze2011oblique} introduced an approach to estimate VI for oblique RFs that computes an analysis of variance (ANOVA) table in non-leaf nodes to obtain p-values for each predictor contributing to the node. The ANOVA VI\footnote{\citet{menze2011oblique} name their method `oblique RF VI', but we use the name `ANOVA VI' in this article to avoid confusing Menze's approach with other approaches to estimate VI for oblique RFs.} is then defined for each predictor as the number of times a p-value associated with the predictor is $\leq 0.01$ while growing a forest. \citet{lundberg2017unified} introduced a method to estimate VI using SHapley Additive exPlanation (SHAP) values, which estimate the contribution of a predictor to a model's prediction for a given observation. SHAP VI is computed for each predictor by taking the mean absolute value of SHAP values for that predictor across all observations in a given set. With the exception of \citet{menze2011oblique}, few studies have evaluated estimation of VI using oblique RFs, and fewer have examined VI specifically for the oblique RSF.

% Several supervised learning algorithms can develop prediction functions for right-censored time-to-event outcomes, henceforth referred to as survival outcomes. \cite{ishwaran2008random} developed the RF for survival, an extension of the RF for regression and classification developed by \citet{breiman2001random}.

\section{The accelerated oblique random survival forest} \label{sec:aorsf}

This section describes our approach to reduce computational overhead of the oblique RSF. Consider the usual framework for right-censored time-to-event outcomes with training data $$\dataset_{\text{train}} = \left\{ (T_i, \delta_i, \bm{x}_{i}) \right\}_{i=1}^{N_{\text{train}}}.$$ Here, $T_i$ is the event time if $\delta_i=1$ or the censoring time if $\delta_i=0$, and $\bm{x}_i$ is a vector of predictors values. Assuming there are no ties, let $t_1 < \, \ldots \, < t_m$ denote the $m$ unique event times in $\dataset_{\text{train}}$.

To accelerate the oblique RSF, we propose to identify linear combinations of predictor variables in non-leaf nodes by applying Newton Raphson scoring to the partial likelihood function of the Cox regression model:
\begin{equation}\label{eqn:cox-partial-likelihood}
L(\bm\beta) = \prod_{i=1}^m \frac{e^{\bm{x}_{j(i)}^T \bm\beta}}{\sum_{j \in R_i} e^{\bm{x}_j^T \bm\beta}},
\end{equation}
where $R_i$ is the set of indices, $j$, with $T_j \geq t_i$ (\ie, those still at risk at time $t_i$), and $j(i)$ is the index of the observation for which an event occurred at time $t_i$. Newton Raphson scoring is an exceptionally fast estimation procedure, and the \texttt{survival} package \citep{survival} includes documentation that outlines how to efficiently program it \citep{therneau_survival_2022}. Briefly, a vector of estimated regression coefficients, $\hat{\beta}$, is updated in each step of the procedure based on its first derivative, $U(\hat{\beta})$, and second derivative, $H(\hat{\beta})$:
$$ \hat{\beta}^{k+1} =  \hat{\beta}^{k} + U(\hat{\beta} = \hat{\beta}^{k})\, H^{-1}(\hat{\beta} = \hat{\beta}^{k}) $$

For statistical inference, it is recommended to continue updating $\hat{\beta}$ by completing additional iterations of Newton Raphson scoring until a convergence threshold is met. However, since an estimate of $\hat{\beta}$ is created by the first iteration of Newton Raphson scoring, only one iteration of Newton Raphson scoring is needed to identify a valid linear combination of predictors. Moreover, computing $U$ and $H$ requires computation and exponentiation of the vector $\bm{x}\hat{\beta}$, but these steps can be skipped on the first iteration of Newton Raphson scoring if an initial value of $\hat{\beta} = 0$ is chosen, allowing for a reduction in computing operations and removing the need to scale predictor values prior to initiating the Newton Raphson algorithm.\footnote{Predictors are scaled prior to initiating the Newton Raphson algorithm to avoid exponentiation of large numbers. However, if only one iteration is completed with an initial value of 0 for $\hat{\beta}$, then $\exp(\bm{x}\hat{\beta}) = 1$.} In \secref{sec:results_pred}, we formally test whether growing oblique survival trees using one iteration of Newton Raphson scoring provides equivalent prediction accuracy compared to trees where iterations are completed until a convergence threshold is met.

Algorithm \ref{alg:aorsf} presents our approach to fitting an oblique survival tree in the accelerated oblique RSF using default values from the \texttt{aorsf} R package. Several steps are taken to reduce computational overhead. First, memory is conserved by conducting bootstrap resampling via randomly generated bootstrap weights rather than making a traditional bootstrap sample. Weights are integer valued, with a weight of $v$ indicating an observation was sampled $v$ times. Second, early stopping is applied to the tree-growing procedure if a statistical criterion is not met. In our case, the criterion is based on the magnitude of a log-rank test statistic corresponding to splitting the data at a current node. Third, instead of greedy recursive partitioning, we use `good enough' partitioning. More specifically, instead of computing a log-rank test statistic for several different linear combinations of variables and proceeding with the highest scoring option, we identify an optimal cut-point for one linear combination of variables and assess whether using this combination will create a split that passes the criterion for splitting a node. If it does not pass the criterion, then another linear combination will be tested, with the maximum number of attempts set by the parameter \texttt{n\_retry}. Often a `good-enough` split can be found in just one attempt when the training set is large, which gives the accelerated oblique RSF a computational advantage in larger training sets compared to greedy partitioning.

\begin{algorithm}
    \caption{Accelerated oblique random survival tree using default parameters.} \label{alg:aorsf}
  \begin{algorithmic}[1]
    \Require Training data $\dataset_{\text{train}} = \left\{ (T_i, \delta_i, \bm{x}_{i}) \right\}_{i=1}^{N_{\text{train}}}$, $\text{mtry} = \sqrt{\text{ncol}(\bm{x}_{\text{train}})}$, $\text{n\_split} = 5$, $\text{n\_retry} = 3$, and $\text{split\_min\_stat} = 3.841459$
    \State $\mathcal{T} \gets \emptyset$
    \State $w \gets \text{sample}(\text{from} = \left\{0, \ldots, 10\right\},\,\text{size} = \text{nrow}(\bm{x}_{\text{train}}),\, \text{replace} = \text{T})$
    \State $\dataset_{\text{in-bag}} \gets \text{subset}(\dataset_{\text{train}},\,\text{rows} = \text{which}(w > 0))$
    \State $w \gets \text{subset}(w,\, \text{which}(w > 0))$
    \State $\text{node\_assignments} \gets \text{rep}(1,\,\text{times} = \text{nrow}(\bm{x}_{\text{in-bag}}))$
    \State $\text{nodes\_to\_split} \gets \{1\}$
     \While {$\text{nodes\_to\_split} \neq \emptyset$}
     \For{$\text{node} \in \text{nodes\_to\_split}$}
       \State $\text{n\_try} \gets 1$
       \State $\text{node\_rows} \gets \text{which}(\text{node\_assignments} \equiv \text{node})$
       \State $\text{node\_cols} \gets \text{sample}(\text{from} = \left\{1, \ldots, \text{ncol}(\bm{x})\right\},\, \text{size} = \text{mtry},\,\text{replace} = \text{F})$ \label{marker}
       \State $\dataset_{\text{node}} \gets \text{subset}(\dataset_{\text{in-bag}},\,\text{rows} = \text{node\_rows},\,\text{columns} = \text{node\_cols})$
       \State $\beta \gets \text{newt\_raph}(\dataset_{\text{node}},\, \text{weights} = \text{subset}(w, \text{node\_rows}),\, \text{max\_iter} = 1)$
       \State $\eta \gets \bm{x}_{\text{node}} \times \beta$
       \State $\mathcal{C} \gets \text{sample}(\text{from} = \text{unique}(\eta),\, \text{size} = \text{n\_split},\,\text{replace} = \text{F})$
       \State $c \gets \argmax_{c^* \in \mathcal{C}} \left\{ \text{log\_rank\_stat}(\eta, c^*) \right\}$
       \If{$\text{log\_rank\_stat}(\eta, c) \geq \text{split\_min\_stat}$}
         \State $\mathcal{T} \gets \text{add\_node}(\mathcal{T},\, \text{name} = \text{node},\, \text{beta} = \beta,\, \text{cutpoint} = c)$
         \State \LeftComment{0}{Right node logic omitted for brevity (identical to left node logic)}
         \State $\text{node\_left\_name} \gets \text{max}(\text{node\_assignments}) + 1$
         \State $\text{node\_left\_rows} \gets \text{subset}(\text{node\_rows},\,\text{which}(\eta \leq c))$
         \State $\text{subset}(\text{node\_assignments}, \text{node\_left\_rows}) \gets \text{node\_left\_name}$
         \If{$\text{is\_splittable}(\text{subset}(\text{node\_assignments}, \text{node\_left\_rows}))$}
           \State $\text{nodes\_to\_split} \gets \text{nodes\_to\_split} \cup \text{node\_left\_name}$
         \Else
           \State $\mathcal{T} \gets \text{add\_leaf}(\mathcal{T},\, \text{data} = \text{subset}(\dataset_{\text{node}},\,\text{rows} = \text{node\_left\_rows}))$
         \EndIf
       \ElsIf{$\text{n\_try} \leq \text{n\_retry}$}
         \State $\text{n\_try} \gets \text{n\_try} + 1$
         \State \Goto{marker}
       \Else
         \State $\mathcal{T} \gets \text{add\_leaf}(\mathcal{T},\, \text{data} =\dataset_{\text{node}})$
       \EndIf
       \State $\text{nodes\_to\_split} \gets \text{nodes\_to\_split} \setminus \{\text{node}\}$
     \EndFor
  \EndWhile
  \State \Return $\mathcal{T}$
  \end{algorithmic}
\end{algorithm}

\section{Negation variable importance} \label{sec:negation_vi}

This Section introduces negation VI, which is similar to permutation VI in that it measures how much a model’s prediction error increases when a variable’s role in the model is de-stabilized. Specifically, negation VI measures the increase in an oblique RF's prediction error after flipping the sign of all coefficients linked to a variable (\ie, negating them). As the magnitude of a coefficient increases, so does the probability that negating it will change the oblique RF's predictions. For the current study, we use Harrell's concordance (C)-statistic \citep{harrell1982evaluating} to measure change in prediction error when computing negation VI.

Negation VI has several helpful characteristics. First, negation VI generalizes to any oblique RF (\ie, not just RSFs) using any valid error function, making it both general and flexible.\footnote{The \texttt{aorsf} package enables customized functions to be applied in lieu of the default C-statistic (see \texttt{?aorsf::orsf\_vi\_negate})} Second, since the coefficients in each non-leaf node of an oblique tree are adjusted for the accompanying predictors, negation VI may provide better estimation of VI in the presence of correlated variables compared to standard VI techniques. Third, unlike permutation, negation is non-random and hence reproducible without setting a random seed. Additionally, since negation VI does not permute variables, the analyst need not worry about impossible combinations of predictors that may occur when one predictor is randomly permuted, such as having a negative status for type 2 diabetes and having Hemoglobin A1c level $\geq$ 6.5\% (a value indicative of type 2 diabetes) as a result of randomly permuting the values of Hemoglobin A1c.

\section{Numeric experiments} \label{sec:numeric}

Sections \ref{sec:bm_pred} and \ref{sec:bm_vi} present numerical experiments examining the accelerated oblique RSF and negation VI, respectively. The code used to run these experiments is available online at \href{https://github.com/bcjaeger/aorsf-bench}{https://github.com/bcjaeger/aorsf-bench}. All analyses were conducted using R version 4.1.3 and coordinated by the \texttt{targets} R package \citep{targets}. To standardize comparisons of computational efficiency, all learners and VI techniques used up to 4 processing units.

\subsection{Benchmark of prediction accuracy and computational efficiency} \label{sec:bm_pred}

The aim of this numeric experiment is to evaluate and compare the accelerated oblique RSF with its predecessor (the oblique RSF from the \texttt{obliqueRSF} R package) and with other machine learning algorithms for risk prediction. Inferences drawn from this experiment include equivalence and inferiority tests based on Bayesian linear mixed models.

\subsubsection{Learners} \label{sec:learners}

We consider four classes of learners: RSFs (both axis-based and oblique), boosting ensembles, regression models, and neural networks. Specific learners from each class are summarized in \tabref{tab:learners}. To facilitate fair comparisons, tuning parameters were harmonized within each class. For example, for RSF learners, we set the minimum node size (a parameter shared by all RSF learners) as 10. Additionally, for RSF learners, the number of randomly selected predictors was the square root of the total number of predictors rounded to the nearest integer, and the number of trees in the ensemble was 500. For boosting, regression, and neural network learners, nested 10-fold cross-validation was applied to tune relevant model parameters. Specifically, tuning for boosting models included identifying the number of steps to complete. For regression models, tuning was used to identify the magnitude of penalization. For neural networks, the number and density of layers was tuned.

\newgeometry{margin=1cm} % modify this if you need even more space
\begin{landscape}

\begin{table}[h!]
\centering
\begin{tabular}{p{2cm} | p{3cm} p{4cm} p{12cm}}
 \hline
 Learner Class & Software & Learners & Description \\ [0.5ex]
 \hline\hline
 \multicolumn{3}{l}{\textit{Random Survival Forests}}\\
 \hline\hline

 Axis based &

 \href{https://www.randomforestsrc.org/index.html}{\texttt{RandomForestSRC}} \newline
 \href{https://CRAN.R-project.org/package=ranger}{\texttt{ranger}} \newline
 \href{http://party.r-forge.r-project.org/}{\texttt{party}} \newline
 \href{https://github.com/whcsu/rotsf}{\texttt{rotsf}} \newline
 \href{https://github.com/whcsu/rsfse}{\texttt{rsfse}} &

 \texttt{rsf-standard} \newline
 \texttt{rsf-extratrees} \newline
 \texttt{cif-standard} \newline
 \texttt{cif-rotate} \newline
 \texttt{cif-spacextend} &

 \texttt{rsf-standard} grows survival trees following Leo Breiman's original random forest algorithm with variables and cut-points selected to maximize a log-rank statistic. \texttt{rsf-extratrees} grows survival trees with randomly selected features and cut-points. \texttt{cif-standard} uses the framework of conditional inference to grow survival trees. \texttt{cif-rotate} extends \texttt{cif-standard} by applying principal component analysis to random subsets of data prior to growing each survival tree. \texttt{cif-spacextend} derives new predictors for each tree in the ensemble, separately. \\ \hline

 Oblique &

 \href{https://CRAN.R-project.org/package=obliqueRSF}{\texttt{obliqueRSF}} \newline
 \href{https://bcjaeger.github.io/aorsf/}{\texttt{aorsf}} &

 \texttt{obliqueRSF-net} \newline
 \texttt{aorsf-net} \newline
 \textcolor{purple}{\texttt{aorsf-fast}} \newline
 \texttt{aorsf-cph} \newline
 \texttt{aorsf-extratrees} &

 Oblique survival trees following Leo Breiman's random forest algorithm. Linear combinations of inputs are derived using \texttt{glmnet} in \texttt{obliqueRSF-net} and \texttt{aorsf-net}, using Newton Raphson scoring for the Cox partial likelihood function in \texttt{aorsf-fast} (1 iteration of scoring) and \texttt{aorsf-cph} (up to 20 iterations), and chosen randomly from a uniform distribution in \texttt{aorsf-extratrees}. Cut-points are selected from 5 randomly selected candidates to maximize a log-rank statistic. \\

 \hline\hline
 \multicolumn{3}{l}{\textit{Boosting ensembles}}\\
 \hline\hline

 Trees &

 \href{https://xgboost.readthedocs.io/en/stable/#}{\texttt{xgboost}} &

 \texttt{xgboost-cox} \newline
 \texttt{xgboost-aft} &

 \texttt{xgboost-cox} maximizes the Cox partial likelihood function, whereas \texttt{xgboost-aft} maximizes the accelerated failure time likelihood function. Nested cross validation (5 folds) is applied to tune the number of trees grown, the minimum number of observations in a leaf node was 10, the maximum depth of trees was 6, and $\sqrt{p}$ variables were considered randomly for each tree split, where $p$ is the total number of predictors. \\

 \hline\hline
 \multicolumn{3}{l}{\textit{Regression models}}\\
 \hline\hline

 Cox Net &

 \texttt{glmnet} &

 \texttt{glmnet-cox} &

 The Cox proportional hazards model is fit using an elastic net penalty. Nested cross validation (5 folds) is applied to tune penalty terms.\\

 \hline\hline
 \multicolumn{3}{l}{\textit{Neural networks}}\\
 \hline\hline

 Cox Time &

 \href{https://raphaels1.github.io/survivalmodels/}{\texttt{survivalmodels}} &

 \texttt{nn-cox} &

 A neural network based on the proportional hazards model with time-varying effects. Nested cross-validation was applied to select the number of layers (from 1 to 8), the number of nodes in each layer (from $\sqrt{p}$/2 to $\sqrt{p}$), and the number of epochs to complete (up to 500). A drop-out rate of 10\% was applied during training.   \\
 \hline

\end{tabular}
\caption{Learning algorithms assessed in numeric studies. \textcolor{purple}{\texttt{aorsf-fast}} is the accelerated oblique random survival forest (see Algorithm \ref{alg:aorsf}), and each of the additional learners are compared to \textcolor{purple}{\texttt{aorsf-fast}} in numeric studies.}
\label{tab:learners}
\end{table}

\end{landscape}
\restoregeometry

\subsubsection{Evaluation of prediction accuracy} \label{sec:prediction_accuracy}

Our primary metric for evaluating the accuracy of predicted risk is the integrated and scaled Brier score \citep{graf1999assessment}, a proper scoring rule that combines discrimination and calibration in one value and improves interpretability by adjusting for a benchmark model \citep{kattan2018index}. Consider a testing data set:
$$\dataset_{\text{test}} = \left\{ (T_i, \delta_i, x_{i}) \right\}_{i=1}^{N_{\text{test}}}.$$
Let $\widehat{S}(t \mid x_i)$ be the predicted probability of survival up to a given prediction time of $t > 0$.
 For observation $i$ in $\dataset_{\text{test}}$, let $\widehat{S}(t \mid \bm{x}_i)$ be the predicted probability of survival up to a given prediction time of $t > 0$. Define \begin{align*}
\bstat = \frac{1}{\ntest} \sum_{i=1}^{\ntest} &\{ \widehat{S}(t \mid \bm{x}_i)^2 \cdot I(T_i \leq t, \delta_i = 1) \cdot \widehat{G}(T_i)^{-1} \\ &+ [1-\widehat{S}(t \mid \bm{x}_i)]^2 \cdot I(T_i > t) \cdot \widehat{G}(t)^{-1}\}
\end{align*} where $\widehat{G}(t)$ is the Kaplan-Meier estimate of the censoring distribution. As $\bstat$ is time dependent, integration over time provides a summary measure of performance over a range of plausible prediction times. The integrated $\bstat$ is defined as \begin{equation}
\bsbar = \frac{1}{t_2 - t_1}\int_{t_1}^{t_2} \widehat{\text{BS}}(t) dt.
\end{equation} In our results, $t_1$ and $t_2$ are the 25th and 75th percentile of event times, respectively. $\bsbar$, a sum of squared prediction errors, can be scaled to produce a measure of explained residual variation (\ie, an $R^2$ statistic) by computing \begin{equation}
R^2 = 1 - \frac{\bsbar}{\bskap}
\end{equation} where $\bskap$ is the integrated Brier score when a Kaplan-Meier estimate for survival based on the training data is used as the survival prediction function $\widehat{S}(t)$. We refer to this $R^2$ statistic as the index of prediction accuracy (IPA) \citep{kattan2018index}.

Our secondary metric for evaluating predicted risk is the time-dependent concordance (C)-statistic. We compute the first time-dependent C-statistic proposed by \citet[][Equation~3]{blanche2013estimating}, which is interpreted as the probability that a risk prediction model will assign higher risk to a case (\ie, an observation with $T \leq t$ and $\delta = 1$) versus a non-case (\ie, an observation with $T > t$). Similar to the IPA, observations with $T \leq t$ and $\delta = 0$ only contribute to inverse probability of censoring weights for the time-dependent C-statistic.

Both the IPA and time-dependent C-statistic generally take values between 0 and 1. To avoid presenting an excessive amount of leading zeroes in our tables, figures, and text, we scale both the IPA and time-dependent C-statistic by 100. For example, we present a value of 25 if the IPA is 0.25, 87 if the time-dependent C-statistic is 0.87, and present 10.2 if the difference between two IPA values is 0.102

\subsubsection{Data sets}

We used a collection of 21 data sets containing a total of 35 risk prediction tasks (tasks per data set ranged from one to four) to benchmark the accelerated oblique RSF versus other learners. Participant-level data from the GUIDE-IT and SPRINT clinical trials and the ARIC, MESA, and JHS community cohort studies was obtained from the National Institute of Health Biologic Specimen and Data Repository Coordinating Center (BioLINCC). Designs and protocols for these studies have been made available \citep{aric1989atherosclerosis, bild2002multi, felker2017effect, sprint2015randomized, taylor2005toward}. All other datasets were publicly available and obtained through R packages (see Appendix A.1). Across all prediction tasks, the number of observations ranged from 137 to 17,549 (median: 1,384), the number of predictors ranged from 7 to 1,692 (median: 41), and the percentage of censored observations ranged from 5.26 to 97.7 (median: 78.1) (\tabrefAppendix{tab:datasets}).

\subsubsection{Monte-Carlo cross validation}

For each risk prediction task, we completed 25 runs of Monte-Carlo cross validation. In each run, we used a random sample containing 50\% of the available data for training and the remaining 50\% for testing each of the learners described in \secref{sec:learners}. Then, for each learner, we computed the IPA, time-dependent C-statistic, and computational time required to fit a prediction model and compute risk predictions. If any learner failed to obtain predictions on any particular split of data\footnote{For example, when the prediction task was to predict risk of death in the ACTG 320 clinical trial (26 events total), some splits did not leave enough events in the training data to fit complex learners such as the neural network}, the results for that split were omitted from downstream analyses.

\subsubsection{Statistical analysis}

After collecting data from 25 replications of Monte-Carlo cross validation for the 14 learners in all 35 risk prediction tasks, we analyzed the resulting 12,250 observations of IPA and, separately, time-dependent C-statistic, using a Bayesian linear mixed model. Our approach follows the ideas described by \citet{benavoli2017time} and \citet{tidymodels}, who developed guidelines on making statistical comparisons between learners using Bayesian models. Specifically, we fit two models: $$\text{IPA} = \widehat{\gamma}_0 + \widehat{\gamma} \cdot \text{learner} + (1\,|\, \text{data/run}) $$ and $$\text{C-stat} = \widehat{\gamma}_0 + \widehat{\gamma} \cdot \text{learner} + (1\,|\, \text{data/run}).$$ Random intercepts for specific splits of data (\ie, \texttt{run} in the model formula) were nested within datasets. The intercept, $\widehat{\gamma}_0$, was the expected value of the outcome using \texttt{aorsf-fast}, making the coefficients in $\widehat{\gamma}$ the expected differences between \texttt{aorsf-fast} and other learners. Default priors from \texttt{rstanarm} were applied for model fitting \citep{rstanarm}.

\paragraph{Hypothesis testing} For both the IPA and time-dependent C-statistic, we conducted equivalence and inferiority tests based on a 1 point region of practical equivalence. More specifically, we concluded that two learners had practically equivalent IPA or time-dependent C-statistic if there was a 95\% or higher posterior probability that the absolute difference in the relevant metric was less than 1. We concluded that one learner was weakly superior when there was $\geq$ 95\% posterior probability that the absolute difference in the relevant metric was non-zero, and concluded superiority when when there was $\geq$ 95\% posterior probability that the absolute difference in the relevant metric was 1 or more.

\subsubsection{Results} \label{sec:results_pred}

A full summary of all results presented in this Section is provided in \tabrefAppendix{tab:bm_pred_all}. In total, 871 out of 875 Monte-Carlo cross validation runs were completed. On run 13, 18, 24 and 25 for the ACTG 320 data, the \texttt{nn-cox} learner encountered an error during its fitting procedure.

\paragraph{Index of prediction accuracy}

Compared to learners that were not oblique RSFs, \texttt{aorsf-fast} had the highest IPA in 20 out of 35 risk prediction tasks, with an overall mean IPA of 12.7 (Figure \ref{fig:bm_pred_viz_ibs}). Compared to the learner with the second highest mean IPA (\texttt{cif-standard}), \texttt{aorsf-fast}'s mean was 1.36 points higher, a relative increase of 12.0\%. The posterior probability of \texttt{aorsf-fast} and \texttt{aorsf-cph} having practically equivalent expected IPA was 0.99, and the posterior probability of \texttt{aorsf-fast} having a superior IPA to other learners ranged from 0.79 (versus \texttt{cif-standard}) to $>$0.999 (versus several other learners; see Figure \ref{fig:bm_pred_model_viz_ibs})

\begin{knitrout}
\definecolor{shadecolor}{rgb}{0.969, 0.969, 0.969}\color{fgcolor}\begin{figure}
\includegraphics[width=\maxwidth]{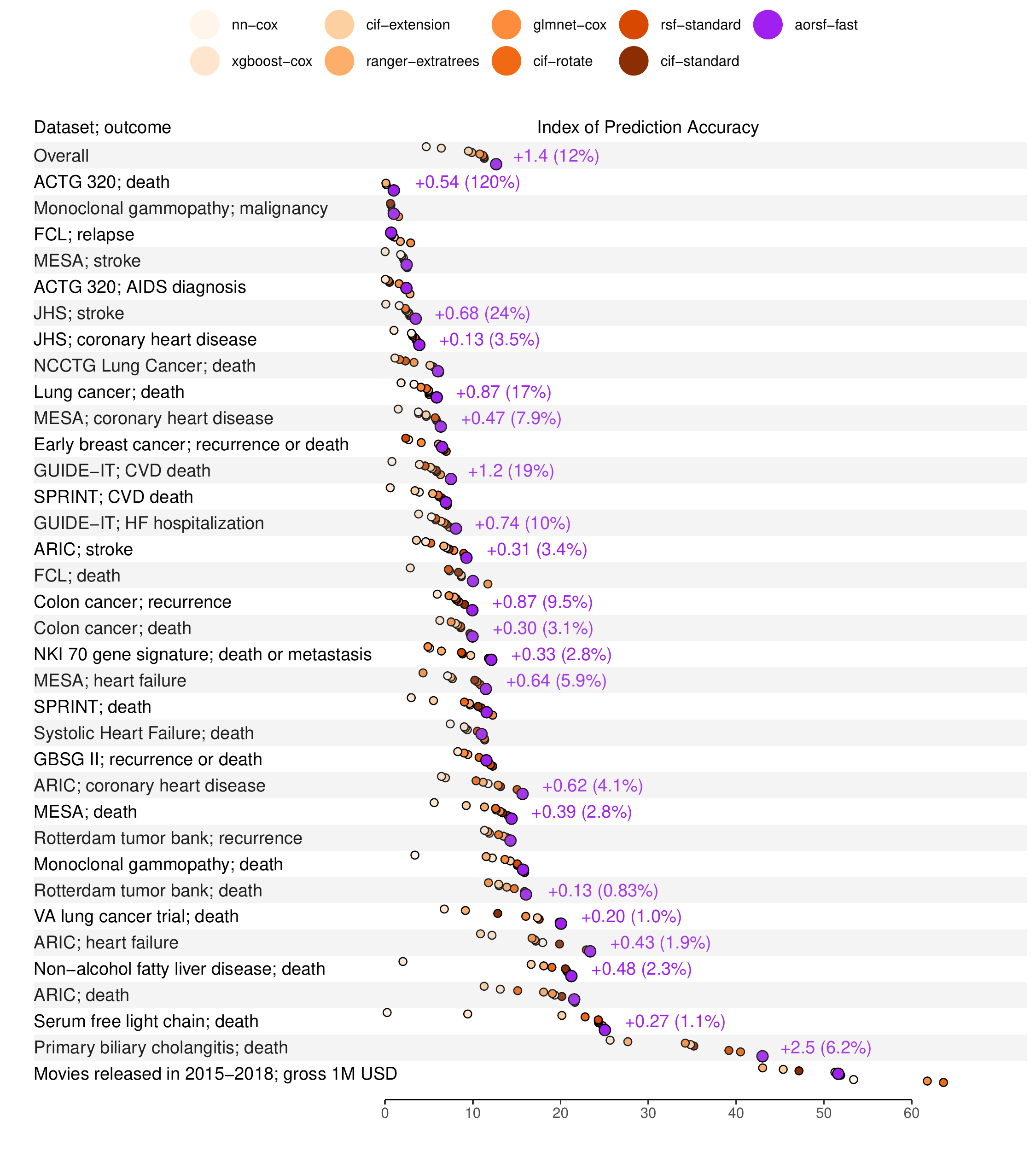} \caption{Index of prediction accuracy for the accelerated oblique random survival forest and other learning algorithms across multiple risk prediction tasks. Text appears in tasks where the accelerated oblique random survival forest obtained the highest index of prediction accuracy, showing the absolute and percent improvement over the second best learner. As predicted survival probabilities are not a standard output from \texttt{xgboost-aft}, it is not included in this figure. Also, since this figure is intended to compare \texttt{aorsf-fast} to learners that are not oblique random survival forests, \texttt{aorsf-cph}, \texttt{aorsf-net}, \texttt{aorsf-random}, and \texttt{obliqueRSF-net} are not included.}\label{fig:bm_pred_viz_ibs}
\end{figure}

\end{knitrout}

\begin{knitrout}
\definecolor{shadecolor}{rgb}{0.969, 0.969, 0.969}\color{fgcolor}\begin{figure}
\includegraphics[width=\maxwidth]{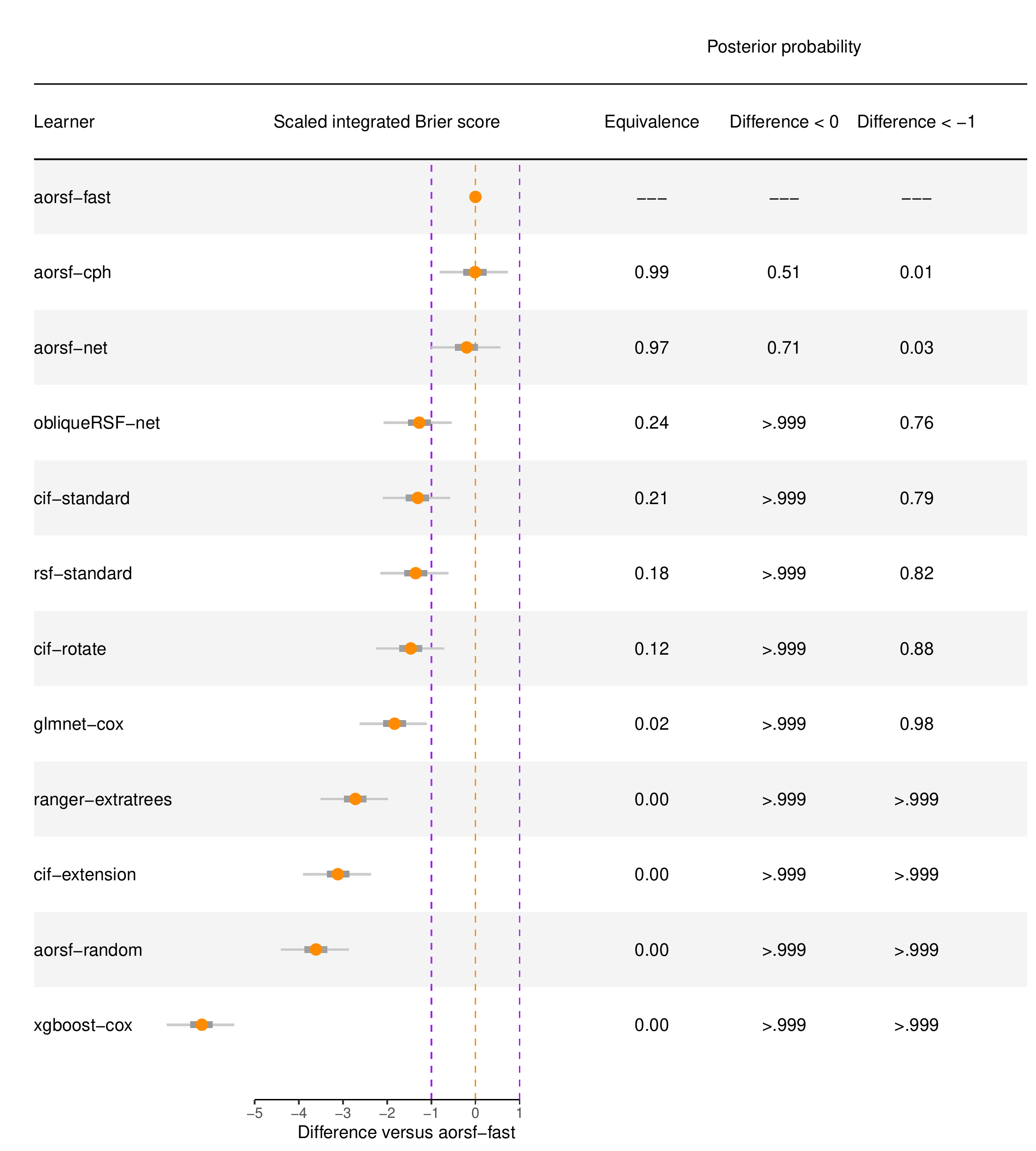} \caption[Expected differences in index of prediction accuracy between the accelerated oblique random survival forest and other learning algorithms]{Expected differences in index of prediction accuracy between the accelerated oblique random survival forest and other learning algorithms. A region of practical equivalence is shown by purple dotted lines, and a boundary of non-zero difference is shown by an orange dotted line at the origin.}\label{fig:bm_pred_model_viz_ibs}
\end{figure}

\end{knitrout}

\paragraph{Time-dependent concordance statistic}

Compared to learners that were not oblique RSFs, \texttt{aorsf-fast} had the highest time-dependent C-statistic in 9 out of 35 risk prediction tasks, with an overall mean of 77.2 (Figure \ref{fig:bm_pred_viz_cstat}). Compared to the learner with the second highest mean C-statistic (\texttt{cif-standard}), \texttt{aorsf-fast}'s mean was 0.721 points higher, a relative increase of 0.943\%. The posterior probability of \texttt{aorsf-fast} and \texttt{aorsf-cph} having practically equivalent expected time-dependent C-statistics was 0.99, and the posterior probability of \texttt{aorsf-fast} having a superior time-dependent C-statistic versus other learners ranged from 0.24 (versus \texttt{cif-standard}) to $>$0.999 (versus several other learners; see Figure \ref{fig:bm_pred_model_viz_cstat})

\begin{knitrout}
\definecolor{shadecolor}{rgb}{0.969, 0.969, 0.969}\color{fgcolor}\begin{figure}
\includegraphics[width=\maxwidth]{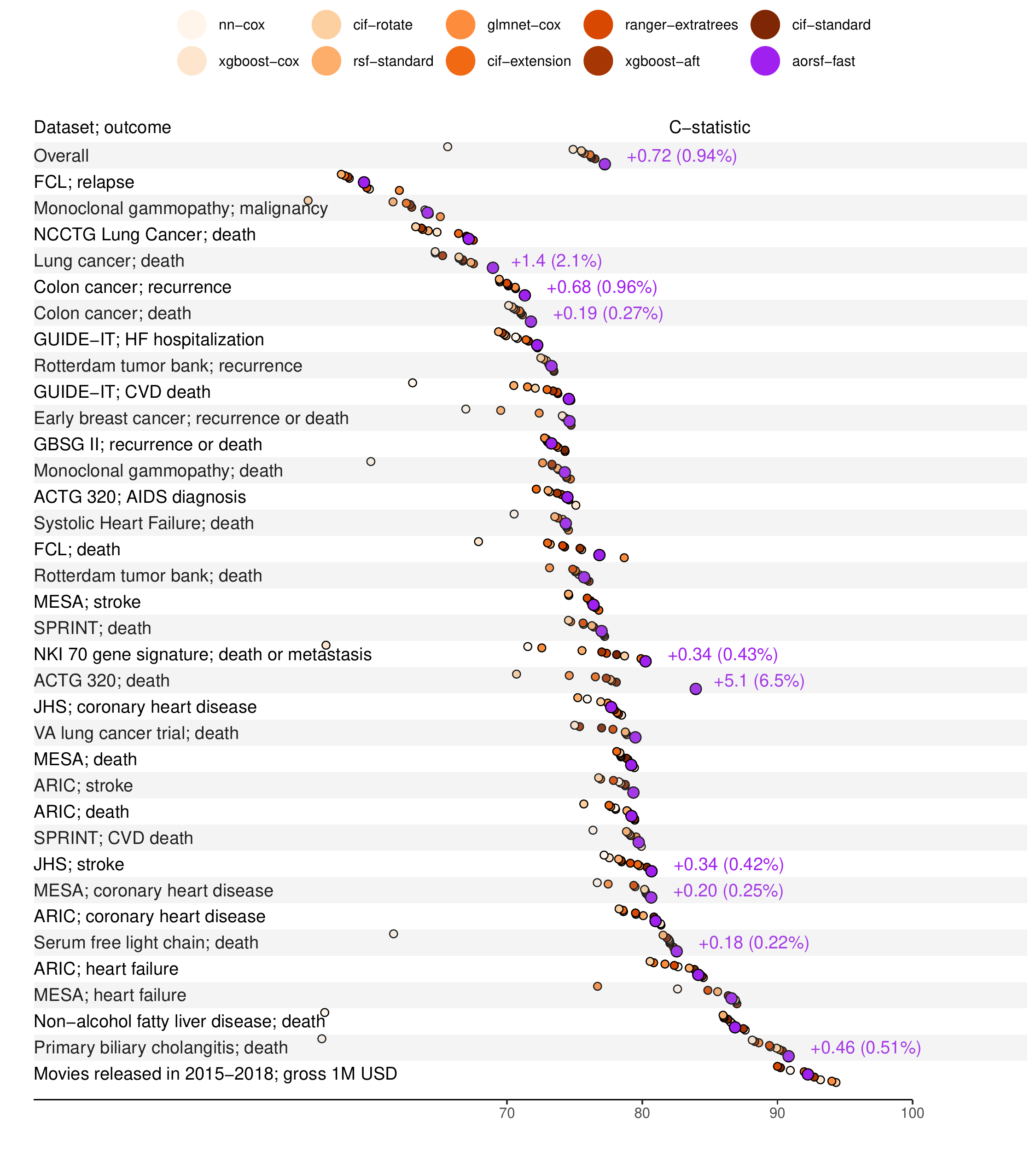} \caption[Time-dependent concordance statistic for the accelerated oblique random survival forest and other learning algorithms across multiple risk prediction tasks]{Time-dependent concordance statistic for the accelerated oblique random survival forest and other learning algorithms across multiple risk prediction tasks. Text appears in tasks where the accelerated oblique random survival forest obtained the highest concordance, showing the absolute and percent improvement over the second best learner. Since this figure is intended to compare \texttt{aorsf-fast} to learners that are not oblique random survival forests, \texttt{aorsf-cph}, \texttt{aorsf-net}, \texttt{aorsf-random}, and \texttt{obliqueRSF-net} are not included.}\label{fig:bm_pred_viz_cstat}
\end{figure}

\end{knitrout}

\begin{knitrout}
\definecolor{shadecolor}{rgb}{0.969, 0.969, 0.969}\color{fgcolor}\begin{figure}
\includegraphics[width=\maxwidth]{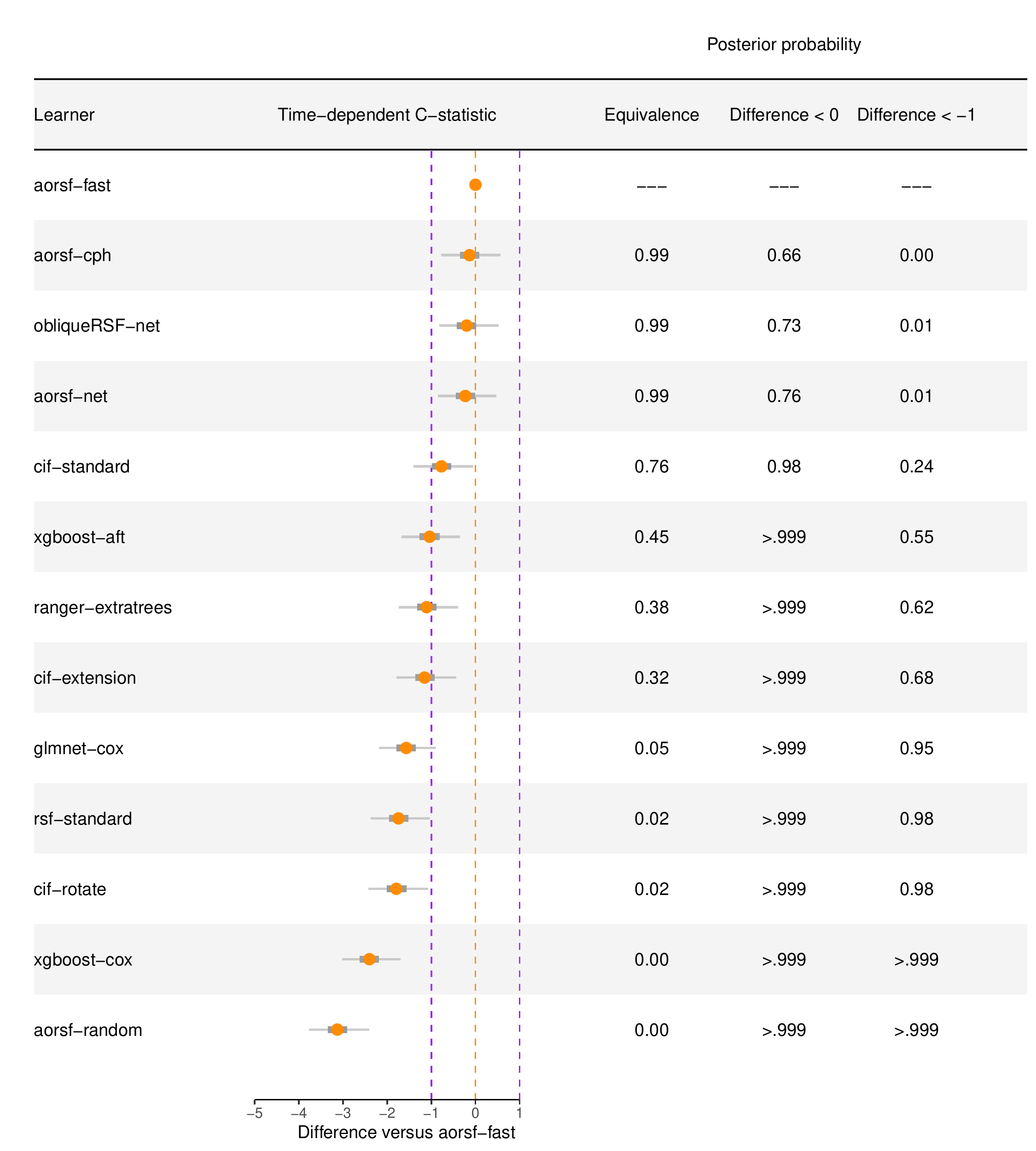} \caption[Expected differences in time-dependent concordance statistic between the accelerated oblique random survival forest and other learning algorithms]{Expected differences in time-dependent concordance statistic between the accelerated oblique random survival forest and other learning algorithms. A region of practical equivalence is shown by purple dotted lines, and a boundary of non-zero difference is shown by an orange dotted line at the origin.}\label{fig:bm_pred_model_viz_cstat}
\end{figure}

\end{knitrout}

\paragraph{Computational efficiency}

Overall, \texttt{aorsf-fast} was the second fastest learner, with an expected model development and risk prediction time about 256 milliseconds longer than \texttt{glmnet-cox} (Figure \ref{fig:bm_pred_time}). Comparing median computing times, \texttt{aorsf-fast} was 446.1 times faster than its predecessor, \texttt{obliqueRSF-net}. In addition, \texttt{aorsf-fast} was 18.9, 1.83, and 3.12 faster than axis based forests grown using the \texttt{party}, \texttt{ranger}, and \texttt{randomForestSRC} packages, respectively.

\begin{knitrout}
\definecolor{shadecolor}{rgb}{0.969, 0.969, 0.969}\color{fgcolor}\begin{figure}
\includegraphics[width=\maxwidth]{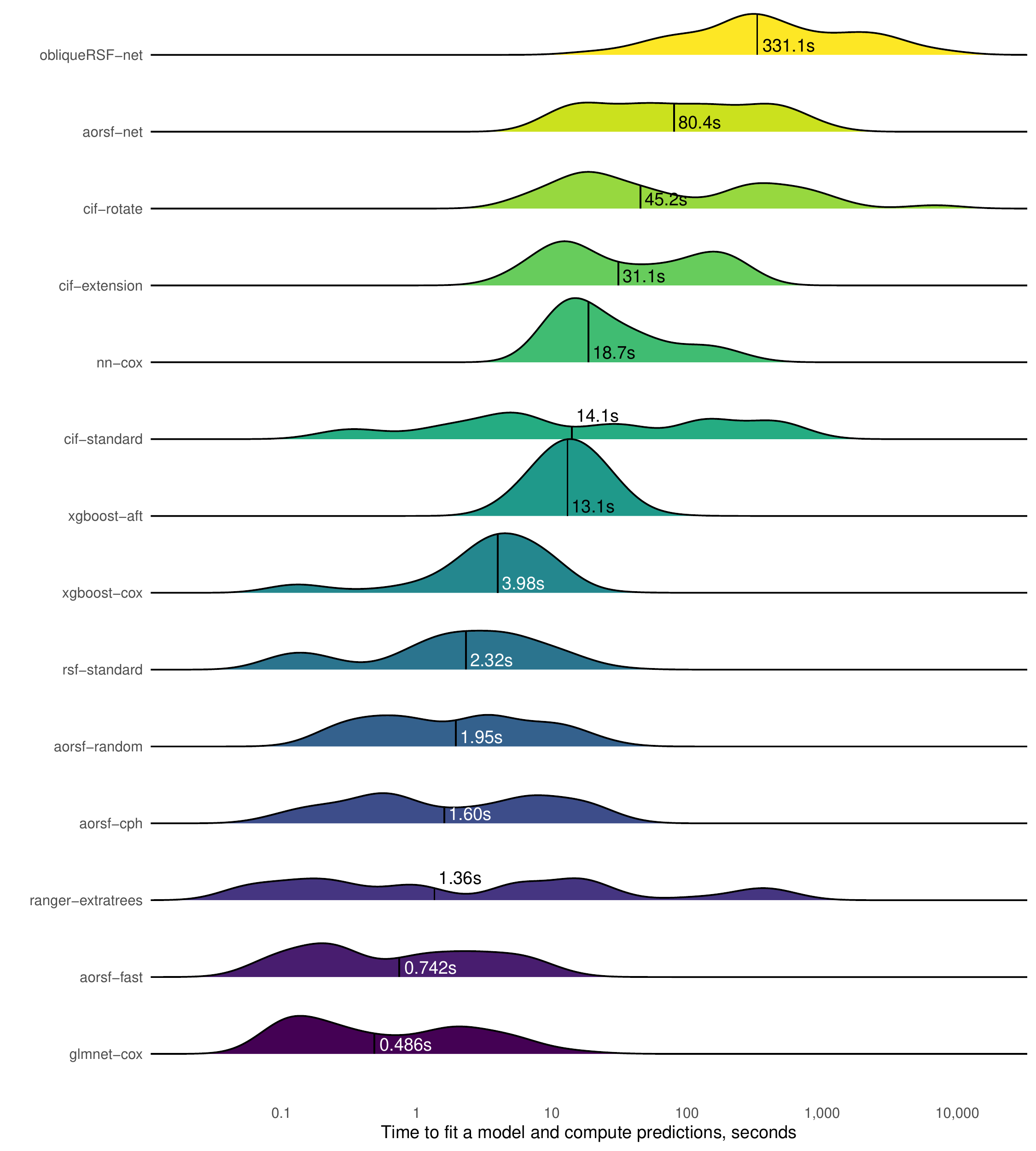} \caption[Distribution of time taken to fit a prediction model and compute predicted risk]{Distribution of time taken to fit a prediction model and compute predicted risk. The median time, in seconds, is printed and annotated for each learner by a vertical line.}\label{fig:bm_pred_time}
\end{figure}

\end{knitrout}

\subsection{Benchmark of variable importance} \label{sec:bm_vi}

The aim of this experiment is to evaluate negation VI and similar VI methods based on how well they can discriminate between relevant and irrelevant variables, where relevance is defined by having a relationship with the simulated outcome. We consider methods that are intrinsic to the oblique RF (\eg, ANOVA VI), those that are intrinsic to the RF (\eg, permutation VI), and those that are model-agnostic (\eg, SHAP VI). VI methods with unavailable or still developing software were not included.\footnote{Although the \texttt{party} package implements the approach to VI developed by \citet{strobl2007bias}, the developers of the \texttt{party} package note that the implementation of this approach for survival outcomes is ``extremely slow and experimental'' as of version 1.3.10. Therefore, it is not incorporated in the current simulation study.}

\subsubsection{Variable importance techniques}

We compute permutation VI for axis based RSFs using the \texttt{randomForestSRC} package. We compute ANOVA VI, negation VI, and permutation VI for oblique RSFs using the \texttt{aorsf} package. For ANOVA VI, we applied a p-value threshold of 0.01, following the threshold recommended by \citet{menze2011oblique}. We compute SHAP VI for boosted tree models using the \texttt{xgboost} package \citep{xgboost}, which incorporates the tree SHAP approach proposed by \citet{lundberg2018consistent}.

% We also compute SHAP VI for accelerated oblique RSFs using the \texttt{fastshap} package.

\subsubsection{Variable types}

We considered five classes of predictor variables, with each class characterized by its variables' relationship to a right-censored outcome. Specifically, \begin{itemize}
\item \textit{irrelevant} variables had no relationship with the outcome.
\item \textit{main effect} variables had a linear relationship to the outcome.
\item \textit{non-linear effect} variables had a non-linear relationship to the outcome.
\item \textit{combination effect} variables were formed by linear combinations of three other variables. While their combination was linearly related to the outcome, each of the three variables contributing to the combination had no relation to the outcome.
\item \textit{interaction effect} variables were related to the outcome by multiplicative interaction with one other variable, which could have been a main effect, non-linear effect, or combination effect variable.
\end{itemize}

\subsubsection{Simulated data}

We initiated each set of simulated data with a random draw of size $n$ from a $p$-dimensional multivariate normal distribution, yielding $n$ observations of $p$ predictors. Each of $p$ predictor variables had a mean of zero, standard deviation of 1, and correlation with other predictor variables drawn at random between a lower and upper boundary. A time-to-event outcome with roughly 45\% of observations censored was generated using the \texttt{simsurv} package. The full predictor matrix (\ie, including interactions, non-linear mappings, and combinations) was used to generate the outcome. Interactions, non-linear mappings, and combinations were dropped from the predictor matrix after the outcome was generated so that VI techniques could be evaluated based on their ability to detect these effects.

\subsubsection{Parameter specifications}

Parameters that varied in the current simulation study included the number of observations (500, 1000, and 2500) and the absolute value of the maximum correlation between predictors (0.3, 0.15, and 0). Parameters that remain fixed throughout the study included the number of predictors in each class (15) and the effect size of each predictor (one standard deviation increase associated with a 64\% increase in relative risk). Using this design for simulated data, the Heller explained relative risk (95\% confidence interval) of our covariates was 88.4 (88.1, 88.7) \citep{heller2012measure} with 2,500 observations.

\subsubsection{Evaluation of variable importance}

We compared VI techniques based on their discrimination (\ie, C-statistic) between relevant and irrelevant variables. Specifically, we generated a binary outcome for each predictor variable based on its relevance (\ie, the binary outcome is 1 if the variable is relevant, 0 otherwise). Treating VI as if it were a ‘prediction’ for these binary outcomes yields a C-statistic which may be interpreted as the probability that the VI technique will rank a relevant variable higher than an irrelevant variable \citep{harrell1982evaluating}.

\subsubsection{Results} \label{sec:results_vi}

The three techniques that used `aorsf' to estimate VI were ranked first (\texttt{aorsf-negate}; $C = 75.9$), second (\texttt{aorsf-anova}; $C = 73.9$), and third (\texttt{aorsf-permute}; $C = 73.2$) in overall mean C-statistic across all of the simulation scenarios, with \texttt{aorsf-negate} obtaining the highest C-statistic in 26 out of 36 VI tasks (Figure \ref{fig:bm_vi_viz}). Among the four relevant variable classes, \texttt{aorsf-negate} had the highest mean C-statistic for main effects, combination effects, and non-linear effects, with the greatest advantage of using \texttt{aorsf-negate} occurring among non-linear and combination variables. Full results from the experiment are provided in \tabrefAppendix{tab:bm_vi}. Computationally, ANOVA VI was faster than negation and permutation VI, with a median time of 2.88 seconds versus 20.4 and 21.8 seconds, respectively.

\begin{knitrout}
\definecolor{shadecolor}{rgb}{0.969, 0.969, 0.969}\color{fgcolor}\begin{figure}
\includegraphics[width=\maxwidth]{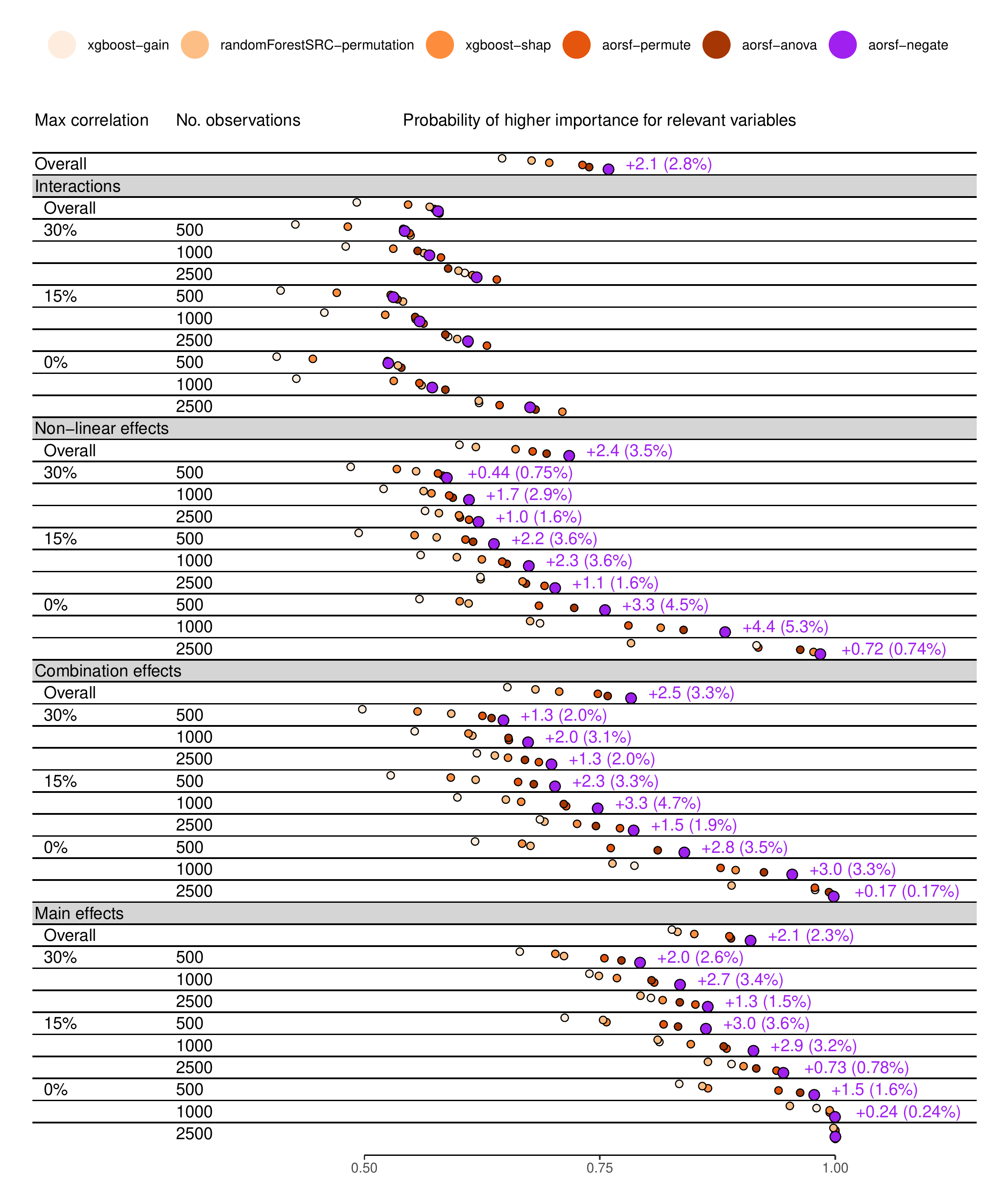} \caption[Concordance statistic for assigning higher importance to relevant versus irrelevant variables]{Concordance statistic for assigning higher importance to relevant versus irrelevant variables. Text appears in rows where negation importance obtained the highest concordance, showing absolute and percent improvement over the second best technique.}\label{fig:bm_vi_viz}
\end{figure}

\end{knitrout}

\section{Discussion} \label{sec:discussion}

In this paper, we have developed two contributions to the oblique RSF: (1) the accelerated oblique RSF (\ie, \texttt{aorsf-fast}) and (2) negation VI. Our technique to accelerate the oblique RSF reduces the number of operations required to find linear combinations of inputs using a single iteration of Newton Raphson scoring, while our VI technique directly engages with coefficients in linear combinations of inputs to measure importance of individual variables. In numeric experiments, we found that that \texttt{aorsf-fast} is approximately 446.1 times faster than its predecessor, \texttt{obliqueRSF-net}, with a practically equivalent C-statistic. We also found that negation VI, a technique to estimate VI using the oblique RSF, detected non-linear, combination, and main effects more effectively than three standard methods to estimate VI: permutation, ANOVA, and SHAP VI. Overall, we found that estimating VI using negation instead of ANOVA increased the C-statistic for ranking a relevant variable higher than an irrelevant variable by 2.05, a relative increase of 2.78\%.

\subsection{Implications of our results}

Accurate risk prediction models have the potential to improve healthcare by directing timely interventions to patients who are most likely to benefit. However, prediction models that cannot scale adequately to large databases or cannot be interpreted and explained will struggle to gain acceptance in clinical practice \citep{moss2022demystifying}. The current study advances the oblique RSF, an accurate risk prediction model, towards being accurate, scalable, and interpretable. The improved computational efficiency of the accelerated oblique RSF increases the feasibility of applying oblique RSFs in a wide range of prediction tasks. Faster model evaluation and re-fitting also improve diagnosis and resolution of model-based issues (\eg, model calibration deteriorates over time). The introduction of negation VI also advances interpretability. VI is intrinsically linked to model fairness, as it can be used to identify when protected characteristics such as race, religion, and sexuality are inadvertently used (either directly or through correlates of these characteristics) by a prediction model. Since negation VI engages with the coefficients used in linear combinations of variables, a major component of oblique RSFs, it may be more capable of diagnosing unfairness in oblique RSFs compared to permutation importance and model-agnostic VI techniques.

\subsection{Limitations and next steps}

While the current study advances the oblique RSF towards being scalable and interpretable, there remain several limitations that can be targeted in future studies. The accelerated oblique RSF does not account for competing risks, and biased estimation of incidence may occur when competing risks are ignored. Thus, allowing the oblique RSF to account for competing risks is a high priority for future studies. In addition, the current study only considered data without missing values, only evaluated oblique RSFs that applied the log-rank statistic for node splitting, and only considered negation VI estimates based on Harrell's C-statistic. Few studies have developed strategies to deal with missing data while growing oblique survival trees. Prior studies have found that log-rank tests can be mis-informative when survival curves cross \citep{li2015statistical}, and that Harrell's C-statistic is dependent on the censoring distribution of the outcome \citep{uno2011c}. Thus, a second item is to expand the range of options available to users of the \texttt{aorsf} package, enabling them to apply strategies for imputation of missing values and use a broad range of statistical criteria while growing oblique survival trees. Last, \citet{cui2017consistency} found that estimating an inverse-probability weighted hazard function at each non-leaf node of a survival tree allows the RSF to converge asymptotically to the true survival function when some variables contribute both to the risk of the event and the risk of censoring, a scenario that is very likely in the analysis of medical data. The accelerated oblique RSF could incorporate this splitting technique by using Newton Raphson scoring to fit a model for the censoring distribution after which a weighted model could be fit to the failure distribution. This final item has the highest priority, as \citet{cui2017consistency} showed it is a requisite condition for consistency of axis-based survival trees in fairly general settings.

% Acknowledgements should go at the end, before appendices and references

Oblique RSFs have exceptional prediction accuracy and this study has shown how they can be fit with computational efficiency that rivals their axis-based counterparts. We have also introduced a general and flexible method to estimate VI with oblique RFs, and demonstrated its effectiveness specifically for the oblique RSF. Code used for the current study is available at \href{https://github.com/bcjaeger/aorsf-bench}{https://github.com/bcjaeger/aorsf-bench}, and the \texttt{aorsf} package is available at \href{https://github.com/bcjaeger/aorsf}{https://github.com/bcjaeger/aorsf}.

\subsection*{Acknowledgements}

Research reported in this publication was supported by the Center for Biomedical Informatics, Wake Forest University School of Medicine. The project described was supported by the National Center for Advancing Translational Sciences (NCATS), National Institutes of Health, through Grant Award Number UL1TR001420. The content is solely the responsibility of the authors and does not necessarily represent the official views of the NIH.


\newpage

\appendix

\section*{Appendix}
\captionsetup{labelformat=AppendixTables}
\setcounter{table}{0}

\subsection*{Data sources}

\begin{enumerate}

 \item The ``VA lung cancer trial'' data \citep{kalbfleisch2011statistical} were obtained from the \texttt{randomForestSRC} R package \citep{randomForestSRC}. 
 \item The ``Colon cancer'' data \citep{moertel1995fluorouracil} were obtained from the \texttt{survival} R package \citep{survival}. 
 \item The ``Primary biliary cholangitis'' data \citep{therneau2000cox} were obtained from the \texttt{aorsf} R package \citep{aorsf}. 
 \item The ``Movies released in 2015-2018'' data  were obtained from the \texttt{censored} R package \citep{censored}. 
 \item The ``GBSG II'' data \citep{schumacher1994rauschecker} were obtained from the \texttt{TH.data} R package \citep{TH.data}. 
 \item The ``Systolic Heart Failure'' data \citep{hsich2011identifying} were obtained from the \texttt{randomForestSRC} R package \citep{randomForestSRC}. 
 \item The ``Serum free light chain'' data \citep{dispenzieri2012use, kyle2006prevalence} were obtained from the \texttt{survival} R package \citep{survival}. 
 \item The ``Non-alcohol fatty liver disease'' data \citep{allen2018nonalcoholic} were obtained from the \texttt{survival} R package \citep{survival}. 
 \item The ``Rotterdam tumor bank'' data \citep{royston2013external} were obtained from the \texttt{survival} R package \citep{survival}. 
 \item The ``ACTG 320'' data \citep{hosmer2002applied} were obtained from the \texttt{mlr3proba} R package \citep{mlr3proba}. 
 \item The ``GUIDE-IT'' data \citep{felker2017effect} were obtained from BioLINCC. 
 \item The ``Early breast cancer'' data \citep{desmedt2011multifactorial, hatzis2011genomic, ternes2017identification} were obtained from the \texttt{biospear} R package \citep{biospear}. 
 \item The ``SPRINT'' data \citep{sprint2015randomized} were obtained from BioLINCC. 
 \item The ``NKI 70 gene signature'' data \citep{van2002gene} were obtained from the \texttt{OpenML} R package \citep{OpenML}. 
 \item The ``Lung cancer'' data \citep{director2008gene} were obtained from the \texttt{OpenML} R package \citep{OpenML}. 
 \item The ``NCCTG Lung Cancer'' data \citep{loprinzi1994prospective} were obtained from the \texttt{survival} R package \citep{survival}. 
 \item The ``FCL'' data \citep{pintilie2006competing} were obtained from the \texttt{randomForestSRC} R package \citep{randomForestSRC}. 
 \item The ``Monoclonal gammopathy'' data \citep{kyle2002long} were obtained from the \texttt{survival} R package \citep{survival}. 
 \item The ``MESA'' data \citep{bild2002multi} were obtained from BioLINCC. 
 \item The ``ARIC'' data \citep{aric1989atherosclerosis} were obtained from BioLINCC. 
 \item The ``JHS'' data \citep{taylor2005toward} were obtained from BioLINCC.

\end{enumerate}

\begin{knitrout}
\definecolor{shadecolor}{rgb}{0.969, 0.969, 0.969}\color{fgcolor}
\begin{landscape}
\begin{longtable}[t]{lcclcc}
\caption{\label{tab:unnamed-chunk-10}Data sets used for numeric experiments \label{tab:datasets}}\\
\toprule
Label & N observations & N predictors & Outcome & N Events & \% Censored\\
\midrule
VA lung cancer trial & 137 & 8 & Death & 128 & 6.57\\
\cmidrule{1-6}\pagebreak[0]
 &  &  & Recurrence & 468 & 49.6\\
\cmidrule{4-6}\nopagebreak
\multirow{-2}{*}{\raggedright\arraybackslash Colon cancer} & \multirow{-2}{*}{\centering\arraybackslash 929} & \multirow{-2}{*}{\centering\arraybackslash 12} & Death & 452 & 51.3\\
\cmidrule{1-6}\pagebreak[0]
Primary biliary cholangitis & 276 & 19 & Death & 111 & 59.8\\
\cmidrule{1-6}\pagebreak[0]
Movies released in 2015-2018 & 551 & 46 & Gross 1M USD & 522 & 5.26\\
\cmidrule{1-6}\pagebreak[0]
GBSG II & 686 & 10 & Recurrence Or Death & 299 & 56.4\\
\bottomrule
\end{longtable}
\end{landscape}

\end{knitrout}

\newpage

\begin{knitrout}
\definecolor{shadecolor}{rgb}{0.969, 0.969, 0.969}\color{fgcolor}
\begin{longtable}[t]{lcccc}
\caption{\label{tab:unnamed-chunk-11}Index of prediction accuracy, time-dependent concordance statistic, and computational time required to fit and compute predictions for several learning algorithms across 35 risk prediction tasks.\label{tab:bm_pred_all}}\\
\toprule
\multicolumn{1}{c}{ } & \multicolumn{2}{c}{Performance metric (SD)} & \multicolumn{2}{c}{Computation time, seconds} \\
\cmidrule(l{3pt}r{3pt}){2-3} \cmidrule(l{3pt}r{3pt}){4-5}
  & Scaled Brier & C-Statistic & Model fitting & Risk prediction\\
\midrule
\endfirsthead
\caption[]{Index of prediction accuracy, time-dependent concordance statistic, and computational time required to fit and compute predictions for several learning algorithms across 35 risk prediction tasks.\label{tab:bm_pred_all} \textit{(continued)}}\\
\toprule
  & Scaled Brier & C-Statistic & Model fitting & Risk prediction\\
\midrule
\endhead

\endfoot
\bottomrule
\endlastfoot
\addlinespace[0.3em]
\multicolumn{5}{l}{\textit{\textbf{Overall}}}\\
\hline
\hspace{1em}aorsf-fast & 0.127 (0.109) & 0.772 (0.071) & 0.593 & 0.146\\
\hspace{1em}aorsf-cph & 0.126 (0.109) & 0.771 (0.070) & 1.344 & 0.148\\
\hspace{1em}aorsf-net & 0.124 (0.112) & 0.771 (0.071) & 79.741 & 0.158\\
\hspace{1em}obliqueRSF-net & 0.113 (0.085) & 0.771 (0.071) & 300.393 & 18.803\\
\hspace{1em}cif-standard & 0.113 (0.098) & 0.765 (0.071) & 4.293 & 6.044\\
\hspace{1em}rsf-standard & 0.113 (0.114) & 0.755 (0.075) & 2.099 & 0.199\\
\hspace{1em}cif-rotate & 0.112 (0.124) & 0.755 (0.081) & 37.695 & 8.348\\
\end{longtable}

\end{knitrout}

\begin{knitrout}
\definecolor{shadecolor}{rgb}{0.969, 0.969, 0.969}\color{fgcolor}
\begin{landscape}
\begin{longtable}[t]{llcccccc}
\caption{\label{tab:unnamed-chunk-12}Discrimination of relevant versus irrelevant variables for several techniques to estimate variable importance. \label{tab:bm_vi}}\\
\toprule
\multicolumn{2}{c}{ } & \multicolumn{3}{c}{accelerated oblique RSF} & \multicolumn{2}{c}{xgboost} & \multicolumn{1}{c}{RSF} \\
\cmidrule(l{3pt}r{3pt}){3-5} \cmidrule(l{3pt}r{3pt}){6-7} \cmidrule(l{3pt}r{3pt}){8-8}
Max correlation & No. observations & Negation & ANOVA & Permutation & SHAP & Gain & Permutation\\
\midrule
\endfirsthead
\caption[]{Discrimination of relevant versus irrelevant variables for several techniques to estimate variable importance. \label{tab:bm_vi} \textit{(continued)}}\\
\toprule
Max correlation & No. observations & Negation & ANOVA & Permutation & SHAP & Gain & Permutation\\
\midrule
\endhead

\endfoot
\bottomrule
\endlastfoot
Overall & Overall & 75.9 & 73.9 & 73.2 & 69.6 & 64.6 & 67.7\\
\addlinespace[0.3em]
\multicolumn{8}{l}{\textit{\textbf{Interactions}}}\\
\hline
\hspace{1em}Overall & Overall & 57.8 & 57.4 & 58.0 & 54.6 & 49.2 & 56.9\\
\hspace{1em}30 & 500 & 54.3 & 54.1 & 54.8 & 48.2 & 42.7 & 54.9\\
\hspace{1em}30 & 1,000 & 56.9 & 55.7 & 58.1 & 53.1 & 48.0 & 56.3\\
\hspace{1em}30 & 2,500 & 61.9 & 58.9 & 64.1 & 61.5 & 60.7 & 60.0\\
\hspace{1em}15 & 500 & 53.1 & 53.5 & 52.8 & 47.1 & 41.1 & 54.1\\
\hspace{1em}15 & 1,000 & 55.9 & 55.4 & 56.3 & 52.2 & 45.8 & 55.4\\
\end{longtable}
\end{landscape}

\end{knitrout}

\vskip 0.2in
\bibliographystyle{unsrtnat}
\bibliography{main}

\end{document}